\documentclass[prd,twocolumn,showpacs,preprintnumbers,amsmath,amssymb]{revtex4}
\usepackage{graphicx}
\begin{document}
%\preprint{DRAFT}

%\setlength{\topmargin}{-0.25in}
%\title{Dyson's hierarchical model with $N$-components and O($N$) symmetry}
\title{The non-perturbative part of the plaquette in quenched QCD}
\author{Y. Meurice}
\email[]{yannick-meurice@uiowa.edu}
\affiliation{Department of Physics and Astronomy\\ The University of Iowa\\
Iowa City, Iowa 52242, USA }
\date{\today}
\begin{abstract}
We define the non-perturbative part of a quantity as the difference between its numerical value and the perturbative series truncated by dropping the order of minimal contribution and the higher orders. For the anharmonic oscillator, the double-well potential and the single plaquette gauge theory, the non-perturbative part can be parametrized as 
$A\ \lambda^B \ {\rm e}^{-C/\lambda}$ and the coefficients can be calculated analytically. 
For lattice QCD in the quenched approximation, the perturbative series for the average 
plaquette is dominated at low order by a singularity in the complex coupling plane and 
the asymptotic behavior can only be reached by using extrapolations of the existing series.  We discuss two extrapolations that provide a consistent description of the series up to order 20-25.
These extrapolations favor the idea that the non-perturbative part scales like $(a/r_0)^4$
with $a/r_0$ defined with the force method. We discuss the large uncertainties associated with this statement. We propose a parametrization of ln($(a/r_0)$) as the 
two-loop universal terms plus a constant and exponential corrections. These corrections 
are consistent with $a_{1-loop}^2$ and play an important role when $\beta<6$. We briefly 
discuss the possibility of calculating them semi-classically at large $\beta$. 

\end{abstract}
\pacs{11.15.-q, 11.15.Ha, 11.15.Me, 12.38.Cy}
\maketitle

\section{Introduction}

The perturbative renormalizability of the 
standard model is a remarkable property that played an important role in establishing its phenomenological prominence. It guarantees that we can calculate the radiative corrections at any order in perturbation theory. However, Dyson's argument \cite{dyson52}
suggests that these perturbative series are divergent and one needs to truncate them in order to get a finite result. A procedure often used for this purpose consists in dropping  the smallest contribution at a given coupling and all the higher order terms. The difference between this truncated series and the numerical value of the quantity calculated can be called the non-perturbative part (NPP). The NNP depends on the truncation procedure and also on the renormalization scheme. However, if the NPP has a simple 
parametrization and if we can calculate it approximately, we will have made a big step 
toward a complete solution of the problem. 

In this article, we discuss the NPP as defined above for a variety of models. Our main goal is to find an accurate formula for the average plaquette in lattice QCD in the quenched approximation.
The paper is organized as follows. In Sec. \ref{sec:general}, we calculate the minimal 
error that can be made at a given coupling, if we assume that for sufficiently small coupling, the error is dominated by the first order dropped. In Sec. \ref{sec:anh}, we show 
that the general estimates of Sec. \ref{sec:general} work well for the anharmonic oscillator.  In Sec. \ref{sec:dw}, 
we show that for the ground state of the double-well potential, the one instanton effect is larger than the error estimated on the basis of the asymptotic behavior of the perturbative series. 
However, for the average of the two lowest energy levels, we obtain a good agreement with the general estimate. 
The case of a $SU(2)$ lattice gauge theory with one plaquette is discussed in Sec. \ref{sec:one}. In that case, the NPP can be calculated exactly and it can be checked that the error formula works well. 
In contrast to what happens for the anharmonic oscillator where the error at a given order has a given sign, 
the examples discussed in Secs. \ref{sec:dw} and \ref{sec:one} show more complicated sign patterns that can 
studied easily because we can calculate the numerical values very accurately. In particular, Figs. \ref{fig:anh}  to \ref{fig:npplog} 
should be understood as a tutorial for the interpretation of Figs. \ref{fig:dilog} and \ref{fig:intmodel}. 

In all the above examples the NPP can be approximated as $A\ \lambda^B \ {\rm e}^{-C/\lambda}$ 
with $\lambda$ a generic notation for the expansion parameter. Could the same type of result be obtained in lattice gauge theory? 
Lattice QCD in the quenched approximation is introduced in Sec. \ref{sec:qqcd}. 
Simple hypotheses (power of the two loop renormalization invariant scale, power of the lattice spacing) for the NPP of the plaquette are compared with the numerical data using 
a series calculated up to order 10 \cite{direnzo2000}. This order is not large enough to 
decide if any of the hypothesis is adequate. Higher order extrapolations are necessary to discriminate among the various possibilities. In Sec. \ref{sec:extraps}, two extrapolations are discussed. One is based on 
the existence of complex singularities \cite{third}, the other on the existence of an infrared renormalon \cite{mueller93,itep,burgio97}. These two extrapolations predict quite 
correctly the 16-th coefficient as it can be obtained from a graph in Ref. \cite{rakow05}.
They both seem consistent with the possibility \cite{rakow2002,rakow05} that the NPP is proportional to the 4-th power of the lattice spacing as parametrized using the force method in Refs. \cite{guagnelli98,necco01}. 
The uncertainties associated with the extrapolations are discussed in Sec. \ref{sec:uncertainties}. 
In Sec. \ref{sec:force}, we discuss an exponential parametrization of the lattice spacing as a function of the inverse coupling. Our result 
is consistent with the possibility of corrections scaling approximately like $a_{1-loop}^2$, in good agreement with a suggestion 
made in Ref. \cite{allton96}. In the conclusions, we discuss the possibility of calculating the corrections semi-classically. 

Some words of caution. The separation between the NPP and the truncated series is a computational 
scheme. We do not claim that the NPP has a distinct physical interpretation. 
In the past, there has been attempts \cite{digiacomo81,burgio97,rakow2002,rakow05} to relate the difference between the average plaquette and its perturbative 
expansion to the gluon condensate used in calculations based on the Operator Product Expansion (OPE) in the continuum. The results presented here can certainly be compared to the lattice calculations in these references, 
however our goal is to understand a particular lattice model studied with a particular perturbative scheme. 
We believe that an accurate parametrization of the NPP defined above as a function of $\beta$, is a preliminary step that needs to be completed before attempting to extract accurate numerical values of continuum quantities such as the gluon condensate.

\section{Minimal Error Estimate}
\label{sec:general}

In the following, we consider quantities $Q$ which admit an asymptotic series of the form
\begin{equation}
Q\sim \sum_{k=0}^{\infty} a_k\lambda ^k \ .
\end{equation}
%with coefficients $a_k$ growing at a factorial rate. 
We assume that the leading growth of the coefficients can be parametrized as follows:  
\begin{equation}
a_k\sim C_1C_2^k\Gamma(k+C_3)
\label{eq:growth}
\end{equation}
This type of behavior is justified generically in Ref. \cite{leguillou90}. In the case of lattice gauge theory, this 
type of behavior is seen explicitly in the limiting case of the one plaquette model discussed in Sec. \ref{sec:one}.
We define $\Delta_k$ as the difference between the numerical value of $Q$ and the series truncated at order $k$
\begin{equation}
\Delta _k (\lambda) \equiv Q(\lambda) - \sum_{l=0}^{k} a_l\lambda ^l
\label{eq:deltak}
\end{equation}

In addition, we will assume that for sufficiently small coupling, $\Delta_k$
is approximately given by the next order contribution 
\begin{equation}
\Delta_k \simeq \lambda ^{k+1} a_{k+1}. 
\label{eq:esterr}
\end{equation}
 
At fixed coupling $\lambda$, $|\Delta _k|$ can then be minimized for $k=k^\star$ with 
\begin{equation}
k^\star \simeq (\lambda |C_2|)^{-1}-C_3 - (1/2) +{\cal O}(\lambda |C_2|) \ .
\label{eq:kstar}
\end{equation}

This expression has been obtained by using the leading term of Sterling formula for the gamma function ($\sqrt{2\pi}z^{z-1/2}{\rm e}^{-z}$). In Eq. (\ref{eq:kstar}), $(\lambda |C_2|)^{-1}$ is the leading term and $-C_3 - (1/2)$ the first correction. 
It assumed that $\lambda$ is small enough to neglect 
the  ${\cal O}(\lambda |C_2|)$ terms which require the higher order terms of Sterling formula. Plugging Eq. (\ref{eq:kstar}) in the expected error, gives the minimal error 
\begin{equation}
Min_k \ |\Delta _k|\simeq \sqrt{2\pi}|C_1| (\lambda |C_2|)^{1/2-C_3}
{\rm e}^{-\frac{1}{|C_2|\lambda}} \ .
\label{eq:minerr}
\end{equation}
The exponential part of this formula is well-known \cite{leguillou90}.

\section{The anaharmonic oscillator}
\label{sec:anh}

\begin{figure}
\includegraphics[width=3.2in]{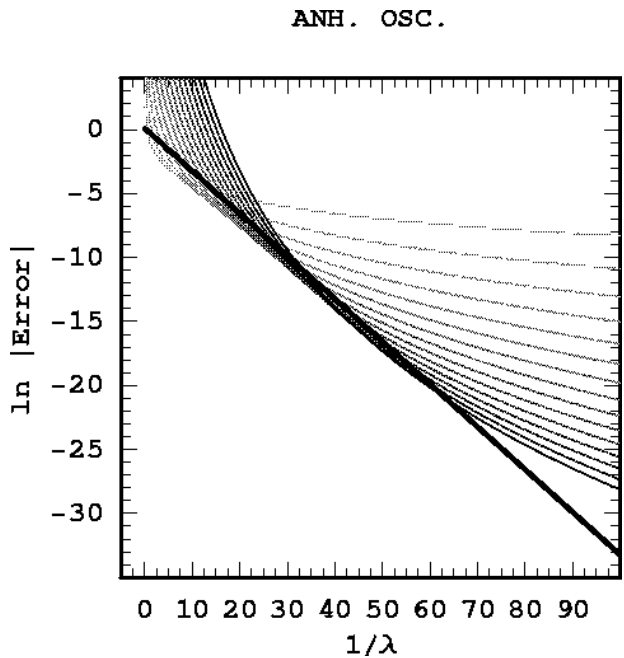}
\caption{Natural logarithm of the absolute value of the difference between the 
series and the numerical value for order 1 to 15 for the anharmonic oscillator as a function of $1/\lambda$. As the order increases, 
the curves get darker. The thicker dark curve is ln ($(\sqrt{12}/\pi){\rm e}^{-\frac{1}{3\lambda}}$) }
\label{fig:anh}
\end{figure}

A simple quantum mechanics example where the situation described in section \ref{sec:general} is approximately realized is the anharmonic oscillator. The Hamiltonian reads 
\begin{equation}
H=p^2/2+ x^2 /2+\lambda x^4 \ .
\end{equation}
We discuss the perturbative expansion of the ground state energy
\begin{equation}
E_0\sim \sum_{k=0}a_k\lambda^k \ .
\end{equation}
The leading asymptotic behavior of the coefficients has been calculated by Bender and Wu \cite{bender69}: 
\begin{equation}
a_k\sim (-1)^{k+1}\sqrt{6/\pi^3}3^k\Gamma(k+1/2)
\end{equation}
Using Eq. (\ref{eq:minerr}), we obtain 
\begin{equation}
Min_k \ |\Delta_k|\simeq (\sqrt{12}/\pi){\rm e}^{-\frac{1}{3\lambda}}\ .
\label{eq:minerranh}
\end{equation}
This is illustrated in Fig. \ref{fig:anh} where the logarithm of the error is plotted 
versus $1/\lambda$. With this choice of variables, the minimal error is a represented by 
a straight line that is approximately tangent to all the curves representing the empirical error. 

\begin{figure}
\includegraphics[width=3.2in]{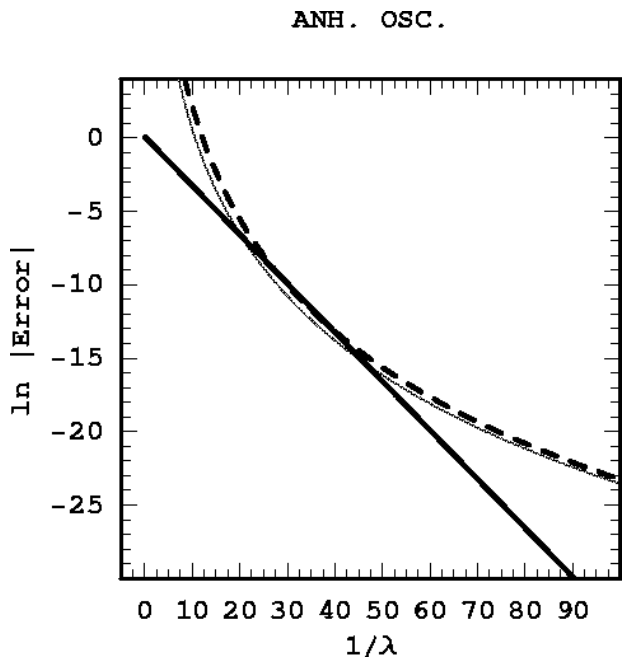}
\caption{Natural logarithm of the absolute value of the difference between the 
series and the numerical value for order 10 as in Fig. \ref{fig:anh} compared with the 
the estimate of Eq. (\ref{eq:esterr}) at order 10 (dashes). The thicker dark curve is ln ($(\sqrt{12}/\pi){\rm e}^{-\frac{1}{3\lambda}}$) }
\label{fig:anh10}
\end{figure}

A careful look at  Fig. \ref{fig:anh} shows that 
the empirical curves go slightly below our estimated error. In other words, $|\Delta_k|$ is slightly smaller than Eq. (\ref{eq:esterr}).
This is illustrated in Fig. \ref{fig:anh10}. One can see that the minimal error from Eq. (\ref{eq:minerranh}) provide an envelope for the curves from Eq. (\ref{eq:esterr}) at various order, but the empirical $|\Delta _k|$ are slightly smaller. One also sees that Eq. (\ref{eq:esterr})
become better as $\lambda$ becomes smaller. A more detailed study shows  that the difference between the Eq. (\ref{eq:esterr}) and $|\Delta _k|$ decreases as a power of $\lambda$ close to 1. The features of $|\Delta _k|$ seem related to the fact that the series has 
alternated signs. In the following, we will only consider same sign series and understanding 
the above feature is not crucial for the rest of the discussion. 


\section{The double-well}
\label{sec:dw}

\begin{figure}
\includegraphics[width=3.2in]{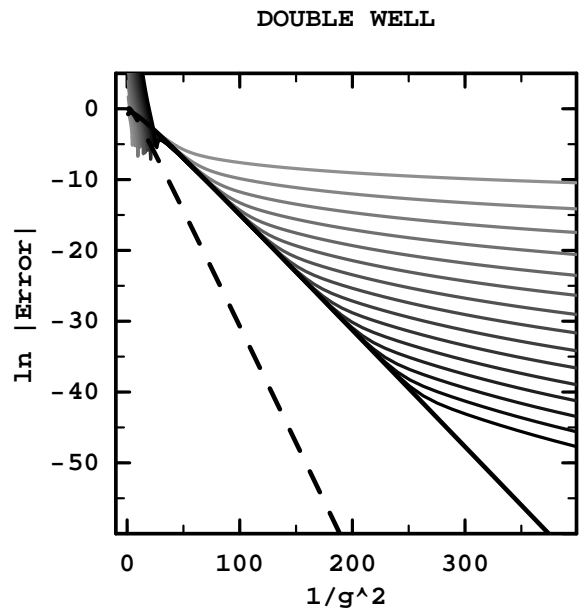}
\vskip20pt
\caption{Natural logarithm of the absolute value of the difference between the 
series and the numerical value for order
 1 to 15 (in $g^2$) versus $1/g^2$ for the ground state of the 
double-well potential. As the order increases, 
the curves get darker.  
The thicker dark curve is ln( $(g\pi)^{-1/2}{\rm e}^{-\frac{1}{6 g^2}}$). 
The dash curve is
ln ($\sqrt{6/\pi}g^{-1}{\rm e}^{-\frac{1}{3g^2}}$).}
\label{fig:dw}
\end{figure}

\begin{figure}
\includegraphics[width=3.2in]{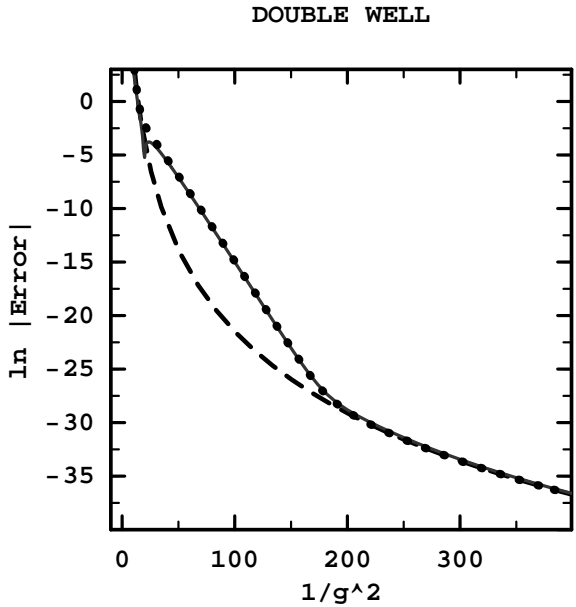}
\caption{Natural logarithm of the absolute value of the difference between the 
series and the numerical value for order
10 (in $g^2$) versus $1/g^2$ for the ground state of the 
double-well potential (solid line). 
The long dash curve is ln ($|a_{11} g^{22}|$).
The dots represent ln ($|-(g\pi)^{-1/2}{\rm e}^{-\frac{1}{6 g^2}}+a_{11} g^{22}|$).
}
\label{fig:dw10}
\end{figure}

A more intricate situation in encountered for the double-well potential. In shifted coordinates, the potential reads
\begin{equation}
V(y)=(1/2)y^2-gy^3+(g^2/2)y^4 \ .
\end{equation}
If we expand the ground state in powers of $g^2$, 
\begin{equation}
E_0\sim \sum_{k=0}a_k\ g^{2k}\ ,
\end{equation}
the leading asymptotic behavior of the coefficients reads \cite{brezin77}
\begin{equation}
a_k\sim - (3/\pi) 3^k\Gamma(k+1) \ .
\label{eq:brezin}
\end{equation}
This implies 
\begin{equation}
Min_k \ |\Delta _k|\simeq \sqrt{6/\pi}g^{-1}{\rm e}^{-\frac{1}{3g^2}}
\label{eq:minerrdw}
\end{equation}
Fig. \ref{fig:dw} shows that this lower bound on the error is not reached. 
Rather, we see that the error is bounded by the one instanton contribution 
\begin{equation}
\Delta E_0 =-(g\pi)^{-1/2}{\rm e}^{-\frac{1}{6 g^2}}\ .
\end{equation}
This is a non-perturbative effect that at first sight is not related 
to the large order behavior of the perturbative series. However, some connection exists (see below).
As illustrated in Fig. \ref{fig:dw10}, a good approximation for the error at order $k$ is 
\begin{equation}
\Delta_k\simeq -(g\pi)^{-1/2}{\rm e}^{-\frac{1}{6 g^2}}+a_{k+1} g^{2(k+1)} \ .
\end{equation}
Both terms are negative and there is no possible cancellation between them.

\begin{figure}
\includegraphics[width=3.2in]{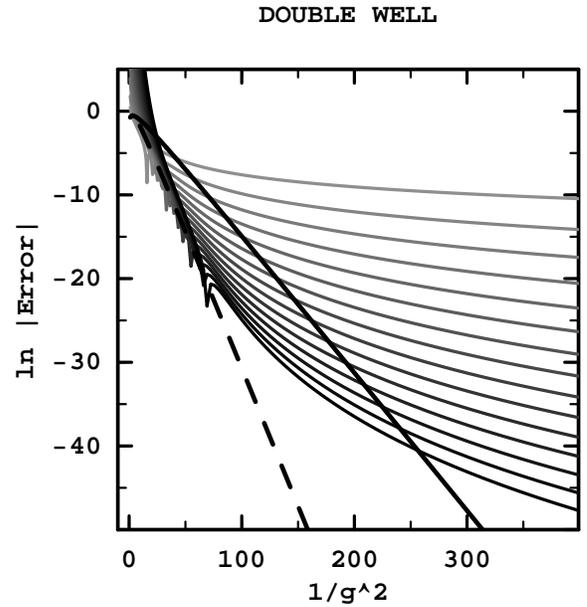}
\caption{Natural logarithm of the absolute value of the difference between the 
series and the numerical value for order
 1 to 10 (in $g^2$) versus $1/g^2$ for the average between the two lowest energy states of the 
double-well potential. As the order increases, 
the curves get darker. 
The thicker dark curve is $(g\pi)^{-1/2}{\rm e}^{-\frac{1}{6 g^2}}$. 
The dash curve is
$\sqrt{6/\pi}g^{-1}{\rm e}^{-\frac{1}{3g^2}}$.}
\label{fig:dwav}
\end{figure}

Except for the overall normalization, Eq. (\ref{eq:minerrdw}) is the square of the 
instanton effect. This is not a coincidence. In Ref. \cite{brezin77}, it is explained that Eq. (\ref{eq:brezin}) 
is an instanton-antiinstanton effect. 
It is possible to get rid of the one instanton effect by 
replacing the numerical value of the ground state by an average over 
the two lowest energy levels. 
This can be seen from the lowest order semi-classical formula \cite{coleman}:
\begin{eqnarray}
E_0&\simeq &(1/2)-(g\pi)^{-1/2}{\rm e}^{-\frac{1}{6 g^2}}\ , \nonumber \\ E_1&\simeq&
(1/2)+(g\pi)^{-1/2}{\rm e}^{-\frac{1}{6 g^2}}\ \nonumber
\end{eqnarray}
One then recovers Eq. (\ref{eq:minerrdw}) as shown in Fig. 
\ref{fig:dwav}. Note that the approximate doubling of the energy levels at small coupling is not seen in perturbation theory because one minimum of the potential goes to infinity when $g$ goes to zero.

\begin{figure}
\includegraphics[width=3.2in]{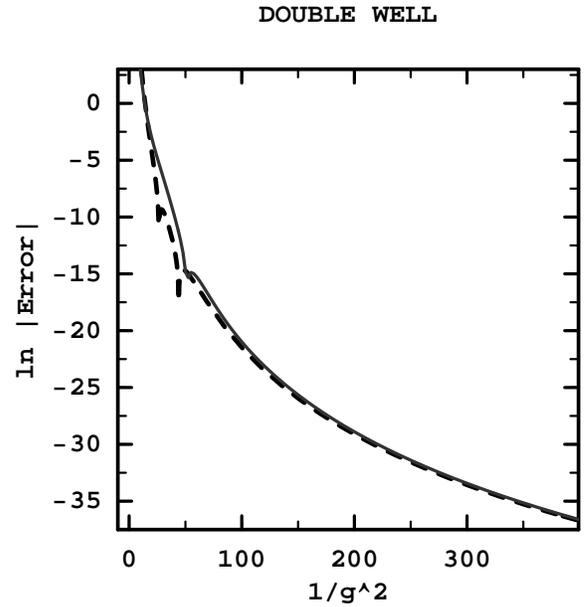}
\caption{Natural logarithm of the absolute value of the difference between the 
series and the numerical value for order
10 (in $g^2$) versus $1/g^2$ for the average of the two lowest states of the 
double-well potential (solid line). 
The dash curve is ln ($|\sqrt{6/\pi}g^{-1}{\rm e}^{-\frac{1}{3g^2}}+a_{11} g^{22}|$)}
\label{fig:dwav5}
\end{figure}

As illustrated in Fig. \ref{fig:dwav5}, a good approximation for the error at order $k$ 
on the average of the two lowest energy states is 
\begin{equation}
\Delta_k^{av.}\simeq \sqrt{6/\pi}g^{-1}{\rm e}^{-\frac{1}{3g^2}}+a_{k+1} g^{2(k+1)} \ .
\end{equation}
The two terms are of opposite signs and there are possible cancellations between them. 
This explains the spikes seen in Figs. \ref{fig:dwav} and \ref{fig:dwav5}.

\section{A one plaquette gauge model}
\label{sec:one}

\begin{figure}
\includegraphics[width=3.2in]{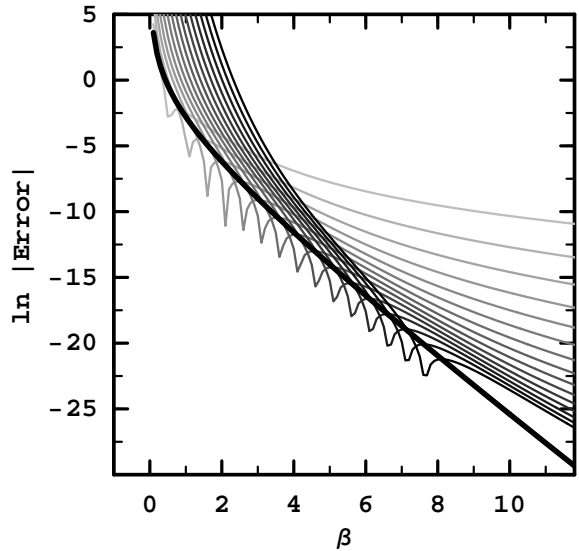}
\caption{Natural logarithm of the absolute value of the difference between the 
series and the numerical value for order 1 to 10 for the one plaquette integral as a function of $\beta$. As the order increases, 
the curves get darker.The thicker dark curve is ln ($(2^{1/2}/\pi)\rm{e}^{-2\beta}/\beta^2$)}
\label{fig:onepl}
\end{figure}

Before discussing quenched QCD, we consider the single plaquette 
$SU(2)$ gauge theory. After gauge fixing three of the four links, the partition function reads
\begin{equation}
Z(\beta)=\int dU {\rm e} ^{-\beta(1-\frac{1}{2}Re TrU)} \ .
\end{equation}
%This can be rewritten as:
%\begin{equation}
%	Z(\beta)=\frac{1}{\pi}\int_0^{2\pi}d\omega \sin^2(\omega/2){\rm e} %^{-\beta(1-\cos(\omega/2))}\ .
%	\label{eq:zsu2}
%\end{equation}
Assuming $\beta>0$, this can be rewritten as 
\begin{equation}
	Z(\beta)=(2/\beta)^{3/2}\frac{1}{\pi}\int_0^{2\beta}dt t^{1/2}
	{\rm e}^{-t}\sqrt{1-(t/2\beta)}
	\label{eq:tint}
\end{equation}
In Ref. \cite{plaquette}, it is shown that the integral can be expressed exactly as a converging expansion in $\beta^{-1}$:
\begin{equation}
\label{eq:notail}
	Z(\beta)=(\beta\pi)^{-3/2} 2^{1/2} 
	\sum_{k=0}^{\infty} A_k(2\beta)\beta^{-k}\ ,
	\end{equation}
	with 
	\begin{equation}
	A_k(x)\equiv 2^{-k}
	\frac{\Gamma(k+1/2)}{k!(1/2-k)}\int_0^{x}dt {\rm e}^{-t}t^{k+1/2}\ ,
	\label{eq:al}
\end{equation}
Note that the peak of the integrand ${\rm e}^{-t}t^{k+1/2}$ becomes larger than $2\beta$ when 
$k= 2\beta-1/2$, consequently the expansion is approximately truncated at that order.
On the other hand, if we extend the range of integration from $2\beta$ to $+\infty$, 
the integral in Eq. (\ref{eq:al}) becomes the gamma function and the coefficients grow 
factorially. 
We can determine the parameters $C_1$, $C_2$ and $C_3$ 
of the leading asymptotic behavior given in Eq. (\ref{eq:growth}). 
We consider the sum  {\it without} the prefactor 
$(\beta\pi)^{-3/2} 2^{1/2}$. From the study of ratios of successive coefficients 
we obtain $C_2=1/2$ and $C_3=0$.   
Using Stirling formula, we then obtain $C_1=1$. 
From Eq. (\ref{eq:kstar}), the optimal order of truncation is 
\begin{equation}
k^\star=2\beta -1/2 \ .
\label{eq:kstar1}
\end{equation}
Note that this is exactly the same order as the order where the peak of the integrand moves 
outside of the range of integration in the exact Eq. (\ref{eq:al}).
From Eq. (\ref{eq:minerr}) and after restoring the prefactor, we obtain 
\begin{equation}
Min_k \ |\Delta _k|\simeq  
(2^{1/2}/\pi)\beta^{-2}\rm{e}^{-2\beta}\ .
\label{eq:mink1}
\end{equation}
Fig. \ref{fig:onepl} shows that this is a good approximation

It is now possible to define precisely what we mean by ``the non-perturbative part of $Z(\beta)$". 
The perturbative part is the perturbative series (Eq. (\ref{eq:al}) with the argument of 
$A_k$ replaced by $\infty$) truncated 
at the integer closest to $k^\star$ defined above, and denoted $r(k^\star)$ hereafter. 
The non-perturbative part consists of two parts. The first part, denoted $R$,
consists in the remaining terms of orders $r(k^\star)+1$ and higher in Eq. (\ref{eq:notail}). The second part, denoted $T$, is minus the integration tails for 
the first $r(k^\star)$ terms that we have added in order to get a perturbative series 
in $1/\beta$ with $\beta$-independent coefficients. More explicitly,
\begin{equation}
Z(\beta)=Z_{Pert.}(\beta)+Z_{NPert.}(\beta)\ ,
\end{equation}
with 
\begin{equation}
Z_{Pert.}(\beta)=(\beta\pi)^{-3/2} 2^{1/2} 
	\sum_{k=0}^{r(k^\star)} A_k(\infty)\beta^{-k}\ ,
\end{equation}
and
\begin{equation}
Z_{NPert.}(\beta)=(R(\beta)-T(\beta))
\end{equation}
\begin{equation}
R(\beta)=(\beta\pi)^{-3/2} 2^{1/2}\sum_{r(k^\star)+1}^{\infty} A_k(2\beta)\beta^{-k}\
\end{equation}
\begin{eqnarray}
&\ & T(\beta)=(\beta\pi)^{-3/2} 2^{1/2}\\ \nonumber
& \times & \sum_{k=0}^{r(k^\star)}\beta^{-k} 
\frac{\Gamma(k+1/2)}{k!(1/2-k)}
\int_{2\beta}^{\infty}dt {\rm e}^{-t}t^{k+1/2} \ ,
\end{eqnarray}
\begin{figure}
\includegraphics[width=3.2in]{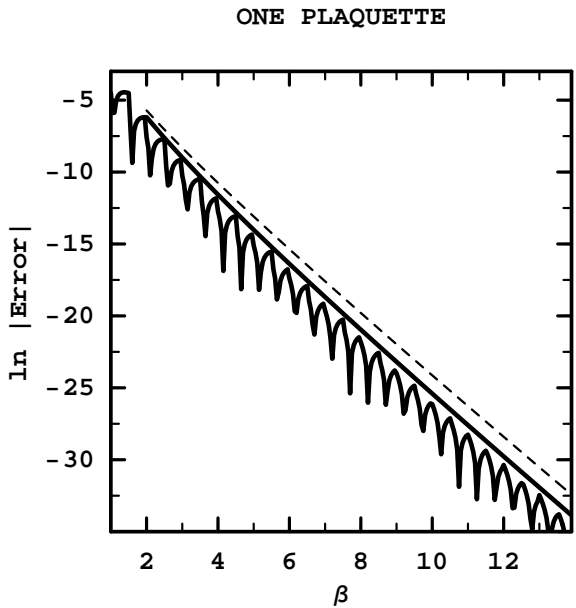}
\caption{ln $|R(\beta)-T(\beta)|$ versus $\beta$ (lowest, wavy line). The intermediate, almost straight curve is ln ($(2^{1/2}/\pi)\beta^{-2}\rm{e}^{-2\beta}$). The dashed curve is ln($  (2/\pi)^{3/2}\beta^{-3/2}\rm{e}^{-2\beta}$)}
\label{fig:nppl}
\end{figure}

The asymptotic behavior of $T(\beta)$ can be estimated using section VI of Ref. \cite{plaquette}.
We obtained that at leading order, the contribution of the tails up to order $K$ is 
\begin{equation}
	\delta Z(\beta, K)\approx A_K{\rm e}^{-2\beta}\beta^{-1}2\pi^{-3/2}\ , 
\end{equation}
with 
\begin{equation}
A_K=-\sum_{l=0}^{K}\frac{\Gamma(l-1/2)}{l!}\ .
\end{equation}
Writing $(1-{\rm e}^{-t})$ as ${\rm e}^{-t}({\rm e}^{t}-1)$ and expanding ${\rm e}^{t}$ we obtain 
\begin{equation}
\sum_{l=1}^{\infty}\frac{\Gamma(l-1/2)}{l!}=\int_0^{\infty}dt t^{-3/2}(1-{\rm e}^{-t})=2\pi^{1/2}\ .
\label{eq:gamma}
\end{equation} 
For large $K$, 
\begin{equation}
\sum_{l=1}^{K}\frac{\Gamma(l-1/2)}{l!}\simeq \int_0^{K}dt t^{-3/2}(1-{\rm e}^{-t})\ ,
\end{equation}
we obtain the leading behavior
\begin{equation}
A_K\simeq\int_K^{\infty}dt t^{-3/2}=2K^{-1/2}
\end{equation}
Replacing $K$ by its optimal value in Eq. (\ref{eq:kstar1}), we conclude that for large $\beta$
\begin{equation}
T(\beta)\simeq (2/\pi)^{3/2}\beta^{-3/2}{\rm e}^{-2\beta}
\end{equation}

\begin{figure}
\includegraphics[width=3.2in]{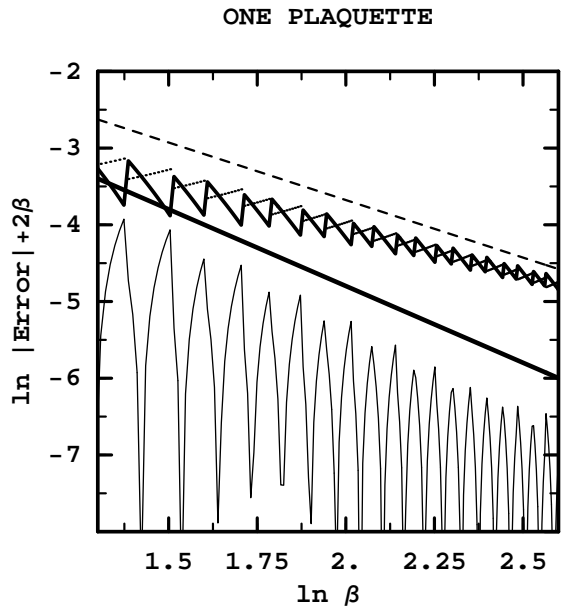}
\caption{ln $|R(\beta)-T(\beta)|+2\beta$ versus ln($\beta$) (wavy curve in the lower part of the graph). The thicker line is ln ($(2^{1/2}/\pi)/\beta^2$). The dashed line is ln($  (2/\pi)^{3/2}\beta^{-3/2}$). ln $|R(\beta)|+2\beta$ (points) and ln $|T(\beta)|+2\beta$ (line) are inter weaved  between the two straight lines. }
\label{fig:npplog}
\end{figure}

It is interesting to notice that $T(\beta)$ is larger by a factor $\beta^{1/2}$ than the 
estimated error given in Eq. (\ref{eq:minerr}). This means that we should also have 
\begin{equation}
R(\beta)\simeq (2/\pi)^{3/2}\beta^{-3/2}{\rm e}^{-2\beta}
\end{equation}
so that the two leading order contributions cancel. The correctness of this argument is illustrated in Figs. 
\ref{fig:nppl} and \ref{fig:npplog}. 

In conclusion, we have learned that in this 
simple lattice model with a compact gauge group, the non perturbative part of the integral has two pieces. 
One piece comes from the higher orders, is positive and dominates at sufficiently large $\beta$. The other comes 
from the added tails of integration, is negative and dominates at sufficiently small $\beta$. It is possible 
that this pattern persists for the usual lattice gauge models on cubic lattices.

\section{Quenched QCD}
\label{sec:qqcd}

We now proceed to introduce lattice quenched QCD, the main model discussed in this article. 
With standard notations, the partition function is 
\begin{equation}
Z=\prod_{l}\int dU_l {\rm e}^{-\beta \sum_{p}(1-(1/N)Re Tr(U_p))} \,
\end{equation}
with $\beta=2N/g^2$.
For symmetric finite lattices with $L^D$ sites and periodic boundary conditions, the number of plaquettes is 
\begin{equation}
\mathcal{N}_p\equiv\ L^D D(D-1)/2\ .
\end{equation}
Using the free energy density
\begin{equation}
f\equiv-(1/\mathcal{N}_p)\ln Z\ ,
\end{equation}
we define the average plaquette
\begin{equation}
	P=\partial f/\partial \beta \ ,
\end{equation}
and its perturbative expansion
\begin{equation}
P(\beta)\sim \sum_{m=1}b_m \beta^{-m} .
\end{equation}
In the following, we used the first three analytical values from Ref. \cite{alles98}, 
and the coefficients from 4 to 10 from Ref. \cite{direnzo2000}. The coefficients of Ref. 
\cite{rakow05} up to order 16 are given in a figure and will be used to check extrapolations. At large $\beta$, high precision data is necessary and we have used 
some values of the average plaquette from Refs. \cite{boyd96,trottier01}.

\begin{figure}
\includegraphics[width=3.2in]{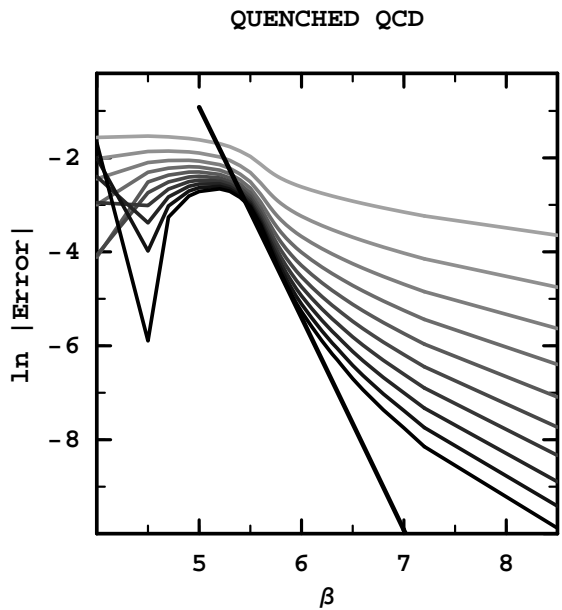}
\caption{Natural logarithm of the absolute value of the difference between the 
series and the numerical value for order 1 to 10 for quenched QCD as a function of $\beta$. As the order increases, 
the curves get darker. The approximately straight dark curve is ln ($6.5 \times 10^8\times(\beta)^{204/121}{\rm e}^{-(16\pi^2/33)\beta}$)}
\label{fig:qcd10}
\end{figure}

\begin{figure}
\includegraphics[width=3.2in]{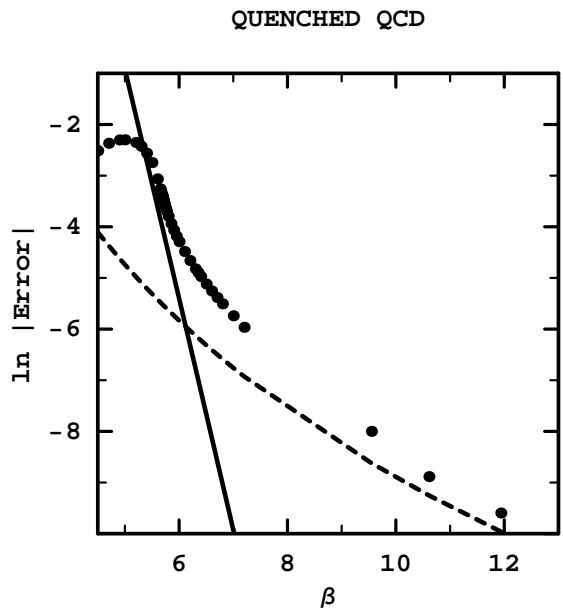}
\caption{Natural logarithm of the absolute value of the difference between the 
series and the numerical value for order 5 for quenched QCD as a function of $\beta$ (dots).
The dash line is ln($|a_6/\beta^6|$).
The solid curve is ln ($6.5 \times 10^8\times(\beta)^{204/121}{\rm e}^{-(16\pi^2/33)\beta}$)}.
\label{fig:qcd5}
\end{figure}

A simple guess is that the envelope of the accuracy curves is given by a power of 
some renormalization group invariant scale.
For instance, one could test the idea that it is proportional to the fourth 
power of the two-loop renormalization 
group invariant scale
\begin{equation}
Min_k \ |\Delta _k|\simeq C (\beta)^{204/121}{\rm e}^{-(16\pi^2/33)\beta} \ .
\label{eq:guessqcd}
\end{equation}
Fig. \ref{fig:qcd10} shows that up to order 10, this provides a reasonable envelope in the region $5.5<\beta<6$. The constant $C$ has been fixed to $6.5 \times 10^8$  by finding a plateau in 
$ |\Delta _{10}|(\beta)^{-204/121}{\rm e}^{(16\pi^2/33)\beta}$.
As the curves leave the conjectured envelope, they become more flat. For instance at order 8 and  
for $6<\beta<7$, a reasonable fit can be obtained \cite{burgio97} with the square of the 
perturbative renormalization invariant scale. However, this does not seem to characterize 
the asymptotic behavior. 
Note also that, at values of $\beta$ close to 6,  
the empirical error is significantly larger than the next order contribution.
This is illustrated in Fig. \ref{fig:qcd5} at order 5.

\begin{figure}
\includegraphics[width=3.2in]{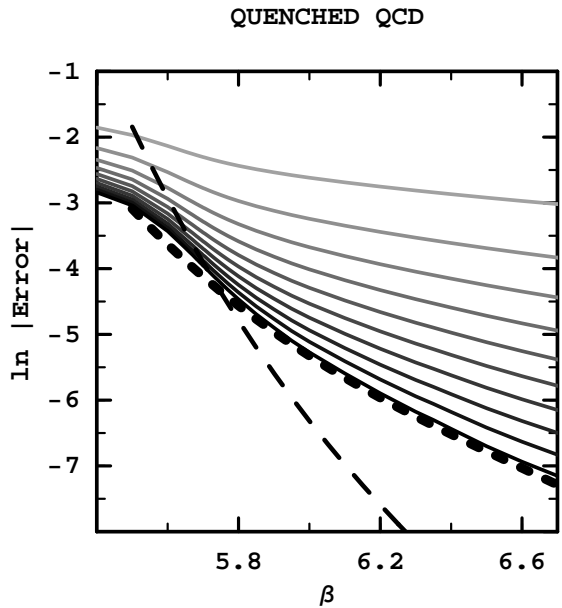}
\caption{Natural logarithm of the absolute value of the difference between the 
series and the numerical value for order 1 to 10 for quenched QCD as a function of $\beta$. As the order increases, 
the curves get darker. The short dash curve is ln($0.14\ (a/r_0)^2$), The long dash curve is ln($1.5\ (a/r_0)^4$).}
\label{fig:qcdsom}
\end{figure}

Another simple guess is that the envelope is proportional to a power of the lattice spacing $a$  expressed in units of $r_0=0.5 fm$. For the interval $5.7<\beta<6.92 $, the following empirical power series \cite{necco01} was obtained from the force method. 
\begin{eqnarray}
{\rm ln}(a/r_0)&=& -1.6804 - 1.7331\ (\beta - 6) \cr \nonumber &+& 0.7849\ (\beta - 6)^2 - 
      0.4428\ (\beta - 6)^3 \ .
\end{eqnarray}
It has been suggested that the envelope is proportional to $a^4$ \cite{digiacomo81,rakow2002,rakow05}, however this is hard to establish without the knowledge of the higher 
orders. On the other hand, the accuracy curve at order 10 can be fitted very well with 
$a^2$ but again it does not seem to characterize the asymptotic behavior since the successive orders are still quite separated. This is illustrated in Fig. \ref{fig:qcdsom}.

The question is now, if the higher orders can be calculated will the envelope stay 
approximately straight 
as in Fig. \ref{fig:qcd10} or curl up as in Fig. \ref{fig:qcdsom}? These two alternatives need to be compared systematically. As 
we will see in the next section, this requires the knowledge of coefficients of higher orders than those 
actually calculated.

\section{Large order extrapolations}
\label{sec:extraps}
In this section we consider two large order extrapolations that can be built out of the 
first 10 terms used above. The first extrapolation is based on the assumption justified 
in Ref. \cite{third} that $\partial P/\partial \beta $ has a logarithmic singularity in the 
complex $\beta$ plane. Integrating, we obtain
\begin{equation}
\sum _{k=1} a_k\beta^{-k}\simeq C({\rm Li}_2 (\beta^{-1}/(\beta_m^{-1}+i\Gamma))+{\rm h.c}\ ,
\label{eq:liser}
\end{equation}
with
\begin{equation}
{\rm Li}_2(x)=\sum _{k=1}x^k/k^2 \ .
\end{equation}
In Ref. \cite{third} we argued that $0.001<\Gamma<0.01$ and for this range of values, the 
low order coefficients depend very little on $\Gamma$. For this reason, it is difficult to fit its value given the
accuracy of the coefficients. We have taken the intermediate value $\Gamma=0.003$ which is compatible with everything we know and determined $C=0.0654$ and $\beta_m$=5.787 using the known values 
of $a_9$ and $a_{10}$ on a $24^4$ lattice \cite{direnzo2000}. 
The estimated errors on these coefficients and their impact on the determination of $C$ and $\beta_m$ are discussed at length in the next section. 
Other choices of $\Gamma$ with this order of magnitude do not affect 
our conclusions. The numerical values are given in Table I. Except for the first term, the agreement is very good. The fact that such a good agreement can be reached by tuning two parameters begs for a diagrammatic explanation!
\begin{table}[t]
\begin{tabular}{||c|c|c||}
\hline
 $m$&$b_m$&$\tilde{b}_m$\cr
 \hline
1 & 2 &  0.7567\cr 
2 & 1.2208 &  1.094 \cr 
    3 & 2.9621 & 2.811\cr 
    4 & 9.417 &  9.138 \cr 
    5 & 34.39 & 33.79\cr 
    6 & 136.8 & 135.5 \cr 
    7 & 577.4 & 575.1\cr 
    8 & 2545 & 2541 \cr 
    9 & 11590 &11590 \cr 
    10 & 54160 & 54160 \cr 
\hline
\end{tabular}
\caption{$b_m$: regular coefficients;  $\tilde{b}_m$: extrapolated coefficients for $\Gamma =0.003$.}
\end{table}
We also obtained $a_{16}=7.7\ 10^8$. This is quite close to what we found from 
Fig. 1 in Ref. \cite{rakow05}:  
$a_{16}\simeq 0.00027\times6^{16}\simeq7.6\ 10^8$. 
Note however, that this value of $a_{16}$ was obtained on a $8^4$ lattice and the finite volume effects 
should be significant. 
The perturbative series obtained  from Eq. (\ref{eq:liser}) has finite radius of convergence \cite{rakow2002,third}.
This seems to contradict the common wisdom \cite{parisi77}  
that if we ignore non-perturbative effects (responsible for the fact that $P$ takes different limits \cite{gluodyn04} when $g^2\rightarrow 0^{\pm}$), we should run into much more serious problems. 
Consequently, we believe that Eq. (\ref{eq:liser}) can only be a good model for the low orders.

Models based on infra-red renormalons \cite{mueller93,itep,burgio97} have a better chance 
to reproduce the large order behavior of the series. We assume
\begin{equation}
P\approx K \int_{t_1}^{t_2}dt {\rm e}^{-\bar{\beta}t}\ (1-t\ 33/16\pi^2)^{-1-x}
\label{eq:renormalon}
\end{equation}
with
\begin{equation}
\bar{\beta}=\beta(1+d_1/\beta+\dots)
\label{eq:shift}
\end{equation}
In Refs. \cite{mueller93,burgio97}, the value $x=204/121$ was used.
Note that $t_1=0$ corresponds to the UV cutoff (we use that value later) and $t_2=16\pi^2/33$ corresponds to the Landau pole.
Expanding $(1-t\ 33/16\pi^2)^{-1-x}$ in powers of $t$ and extending the integration range to $\infty$, we find that at leading order 
\begin{equation}
P(\beta)\sim \sum_{m=1}\bar{b}_m \bar{\beta}^{-m} \ ,
\end{equation}
with 
\begin{equation}
\bar{b}_k=C_1\ (33/16\pi^2)^k\ \Gamma[k+x]\ .
\end{equation}
Setting $d_2$ and higher order coefficients to zero, $x=204/121$ and using the known values of $a_9$ and $a_{10}$ , we obtain $C_1= 0.219$ and $d_1=-3.82$. The low order predictions 
are less good than for the previous method, for instance, $a_6=120.4$. At larger order we obtained  $a_{16}=7.4\ 10^8$ which is close to $7.6\ 10^8$. Note that ${\rm e}^{4 \pi^2 (3.82)/33}\simeq 97$ is significantly larger than the usual \cite{dashen80} factor 28.81 found for the conversion to the 
$\overline{\mbox{MS}}$ scheme.

\begin{figure}
\includegraphics[width=3.2in]{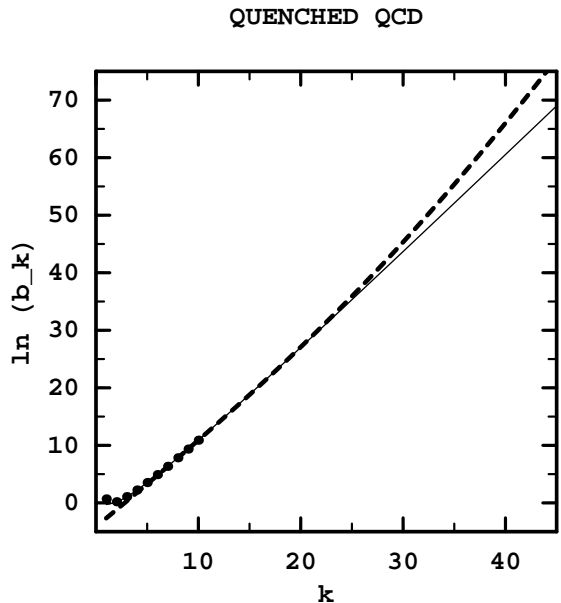}
\caption{ln( $b_k$) for the dilogarithm model (solid line) and the integral model (dashes).
The dots up to order 10 are the known values.}
\label{fig:models}
\end{figure}

The two extrapolations models are compared in Fig. \ref{fig:models}.
The two models yield similar coefficients up to order 20. After that, the integral model 
has the logarithm of its coefficients growing faster than linear.

Using Eq. (\ref{eq:minerr}), for the $\bar{\beta}^{-1}$ expansion, we obtain
\begin{equation}
Min_k \ |\Delta _k|\simeq 3.5(\bar{\beta})^{204/121-1/2}{\rm e}^{-(16\pi^2/33)\bar{\beta}} \ .
\label{eq:guessintmodelbar}
\end{equation}
Shifting to $\beta$ using Eq. (\ref{eq:shift}) and neglecting $\beta^{-1}$ corrections
\begin{equation}
Min_k \ |\Delta _k|\simeq 3.1 \ 10^8(\beta)^{204/121-1/2}{\rm e}^{-(16\pi^2/33)\beta} \ .
\label{eq:guessintmodel}
\end{equation}
Except for the -1/2 in the exponent, this is proportional to the two-loop renormalization group invariant.
The situation is reminiscent of what we encountered in Sec. \ref{sec:one} where the 
error was smaller by a factor $\beta^{-1/2}$ than naively expected. Another possibility 
would be to take $x=204/121+1/2$ if we insist on having a quantity that is scheme independent. 

The study of the asymptotic behavior of the empirical $\beta^{-1}$ expansion between orders 40 and 60 indicates $C_2\simeq 0.203$ quite close to $33/(16\pi^2)\simeq 0.209$. The other values obtained are quite different from those of the coefficients of the $\bar{\beta}^{-1}$ expansion: $C_1\simeq 990$ and $C_3\simeq 4.4$. With these values, 
the optimal order for $\beta=6$ is 25.

\begin{figure}
\includegraphics[width=3.2in]{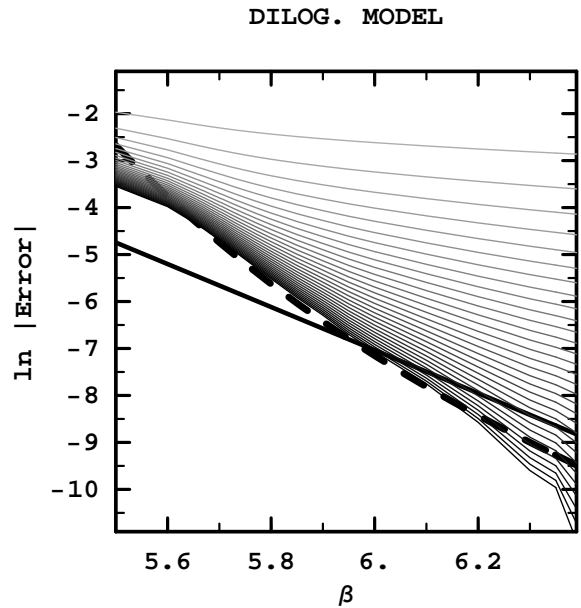}
\caption{Natural logarithm of the absolute value of the difference between the 
series and the numerical value for order 1 to 30 for quenched QCD with the dilogarithm model. As the order increases, 
the curves get darker. The long dash curve is ln($0.65\ (a/r_0)^4$). The solid curve is ln ($3.1\times 10^8\times(\beta)^{204/121-1/2}{\rm e}^{-(16\pi^2/33)\beta}$)}
\label{fig:dilog}
\end{figure}

\begin{figure}
\includegraphics[width=3.2in]{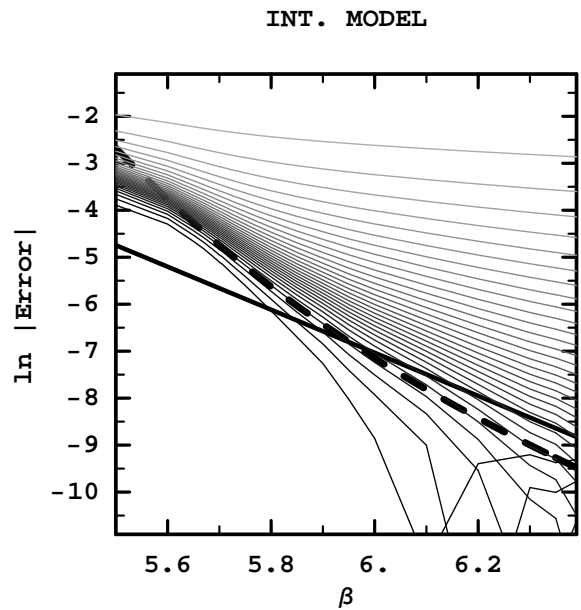}
\caption{Natural logarithm of the absolute value of the difference between the 
series and the numerical value for order 1 to 30 for quenched QCD with the integral model. As the order increases, 
the curves get darker. The long dash curve is ln($0.65\ (a/r_0)^4$). The solid curve is ln ($3.1 \times 10^8\times(\beta)^{204/121-1/2}{\rm e}^{-(16\pi^2/33)\beta}$)}
\label{fig:intmodel}
\end{figure}

The accuracy curves for the two models are compared in Figs. \ref{fig:dilog} and \ref{fig:intmodel}. We have used the known values for the first 10 coefficients. 
A careful examination shows that the figures hardly differ up to order 25. From these two figures, it is plausible that the envelope of the true series is close to $0.65\ (a/r_0)^4$. 
This value of 0.65 does not correspond to a best fit procedure.  If we had forced the $(a/r_0)^4$ behavior and fitted the proportionality constant at order 26, we would have obtained 0.79 for the dilogarithm model and 0.71 
for the integral model. If we had plotted the corresponding curves in Figs \ref{fig:dilog} and 
\ref{fig:intmodel}, they would hardly be visible on these black and white figures. By taking a slightly smaller value 0.65, we made the dash curve more easy to see. We could also have plotted the curve corresponding 
$0.4\ (a/r_0)^4$ that can be inferred from the graphs and the estimate of the gluon condensate of Ref. \cite{rakow05}. Such a curve would be -0.49 below the one we drew.

The tadpole improvement\cite{lepage92} could be another method to study the asymptotic envelope of the accuracy curves.
One defines the new series
\begin{equation}
P \simeq \sum_{m=0}^{K} b_m \beta^{-m}=\sum_{m=0}^{K} e_m \beta_R^{-m} + O(\beta_R^{-K-1})
\end{equation}
with a new expansion parameter
\begin{equation}
\beta_R^{-1}=\beta^{-1}\frac{1}{1-\sum_{m=0} b_m \beta^{-m}} \ .
\end{equation}
The consistent sets of numerical values \cite{third,rakow05,dublin}, show that the beginning 
of the series converges much faster. This is clear up to order 8. However, for larger orders, 
the dependence on the higher order coefficients of the original series due to powers of $1/( 1-\sum_{m=0} b_m \beta^{-m})$ at low orders in $\beta_R$ makes the tadpole improved series 
almost as slow as the original series. In extracting the coefficient \cite{rakow05} of $a^4$, P. Rakow used 
a more sophisticated method where $1/( 1-\sum_{m=0} b_m \beta^{-m})$ is replaced by the 
improved series with a value of $\beta_R$ determined numerically \cite{paulrakow}. 
This procedure gives better perturbative estimates which explains that the constant of proportionality 0.4 is smaller than 0.65 as discussed above.

Instead of assuming the $(a/r_0)^4$ behavior, it is also possible to extract the power from linear fits of ln($|\Delta_k|$) versus ln($a/r_0$).
For $k\leq 26$ the two extrapolations give similar fits near $\beta =6$. For instance, $\Delta_{26}$, 
is $0.56(a/r_0)^{3.72}$ in the dilogarithm model and $0.63(a/r_0)^{3.90}$ for the integral model. For larger $k$, the constant of proportionality and the exponent keep increasing slowly for the dilogarithm model and  the procedure becomes meaningless for the integral model. We now discuss into more detail the stability of this fitting procedure.

\section{Uncertainties on extrapolations}
\label{sec:uncertainties}

We now discuss the effects of the uncertainties of the known coefficients on the extrapolations introduced in the previous section. 
The determination of the unknown parameters in both models depends only on $a_9$ and $a_{10}$. 
The errors on these coefficients are due to the finite volume and the numerical implementation of the stochastic 
perturbation theory. According to Ref. \cite{direnzo2000}, from which we took the values of $a_9$ and $a_{10}$ for  a $24^4$ lattice, the statistical errors on $a_9$ and $a_{10}$ are close to 
one percent while the errors due to the finite volume are less than half a percent for the calculations on a $24^4$ lattice.
 
We have studied the effects of small changes of the following form
\begin{equation}
a_i \rightarrow a_i(1+\delta_i) , 
\end{equation}
for $i=$ 9 and 10 on the determination of the parameters $\cal A$ (amplitude) and $\cal S$ (slope in a log-log plot) appearing in 
the approximate parametrization 
\begin{equation}
\Delta _{26} \simeq {\cal A} (a/r_0)^{\cal S}\ , 
\end{equation}
which was our best estimate of the NPP of the plaquette in the previous section. 
With the sign convention for $\Delta_k$ in Eq. (\ref{eq:deltak}), i. e., the numerical value {\it minus} the series at order $k$, 
$\Delta_k >0$ for  $k\leq 10$. If the coefficients of the series are all positive and end up growing factorially, it is clear that for a given $\beta$, there is some order for which the truncated series becomes larger than the numerical value and $\Delta _k$ becomes negative. This explains the downward spikes that are clearly visible for orders 29 and 30 of the integral model on Fig. \ref{fig:intmodel}. On the other hand, for the dilogarithm model, $\Delta_k$ remains positive as $k$ 
increases, over the whole $\beta$ interval for which $\Delta _k$ is larger than the errors on the numerical values of $P$.  

When the value of the series becomes close to the numerical value, the errors on the Monte Carlo estimates of $P$ become important. The numerical values that we have used for the fits of $\cal A$ and $\cal S$ have estimated errors \cite{boyd96} smaller than $5\times 10^{-5}$. Consequently, the details of Figs. \ref{fig:dilog} and \ref{fig:intmodel} below -10 on the $y$ axis are meaningless. 

We have performed fits of $\cal A$ and $\cal S$ using ${\rm ln}(|\Delta_{26}|)$ versus  ${\rm ln}(a/r_0)$ for 18 values of $\beta$ between 5.68 and 6.10. The quality of the fit can be assessed from 
\begin{equation}
\chi_f\equiv \left(  (1/18)\sum_{i=1}^{18}(D_i/{\rm ln}(|\Delta_{26}(\beta_i)|))^2\right) ^{1/2} \ ,
\end{equation}
with
\begin{equation}
D_i={\rm ln}(|\Delta_{26}(\beta_i)|)-{\rm ln}({\cal A})-{\cal S}{\rm ln}(a(\beta_i)/r_0) \ .
\end{equation}
$\chi_f$ give an estimate of the typical relative errors obtained with the fit. 

It is also useful to assess how well a particular choice of parameters reproduces the known values of the 
known coefficients  of order 8 and less. For this purpose we have defined
\begin{equation}
\chi_a\equiv \left( (1/4)\sum_{j=5}^8[(a_j-a_j[a_9,a_{10}])/a_j]^2\right) ^{1/2} \ ,
\end{equation}
with $a_j[a_9,a_{10}]$ understood as the value of $a_j$ obtained from the models after the unknown coefficients 
have been fixed using $a_9$ and $a_{10}$. We started at order 5 because the differences on lower orders are quite large for the integral model. 

It is interesting to notice that in both models, the extrapolated coefficients depend linearly on a overall scale factor called $C$ in the dilogarithm model and $C_1$ in the integral model. These scale factors cancel if we consider the ratio of the two equations determining the unknown parameters. In other words, the parameters $\beta_m$ and $d_1$ depend only on $a_9/a_{10}$. Consequently if $\delta_9=\delta_{10}=\delta$, $a_9/a_{10}$ is 
unchanged and the only effect of the change is to rescale all the extrapolated coefficients by a common factor $1+\delta$. This type of modification generates a linear response which is the same at all order and has quite controllable effects. 
On the other hand, if $\delta_9\neq \delta_{10}$, the parameters $\beta_m$ and $d_1$ are modified and this has more drastic effects on the extrapolated coefficients.

In order to give a quantitative idea, we have considered the four cases  corresponding to the condition  $|\delta_9|=|\delta_{10}|=0.01$. The results are shown in Tables II and III. The choice $\delta_9=-0.01$  and $\delta_{10}=0.01$ has a $\chi_f$ significantly larger than the other choices. For this choice, 
the two models start differing significantly at order lower than 26
and then develop negative spikes at larger order. 
It might have been a better choice to quote fits at lower order where there is a better agreement between the two models and smaller $\chi_f$. For instance, at $k=22$, we obtain $\cal A$ = 0.69 and $\cal S$ = 3.90 with $\chi_a$ = 0.007 for the dilogarithm model and $\cal A$ = 0.61 and $\cal S$ = 3.77 with $\chi_a$ = 0.005 for the integral  model. As the fit for $k=26$ did not look reasonable for this choice of $\delta_i$, we also added a line with the 
same signs but a lower magnitude.
\begin{table}[t]
\begin{tabular}{||c|c|c|c|c|c|c|c||}
\hline
$\delta_9$&$\delta_{10}$&$C$&$\beta_m$&$\cal A$&$\cal S $ &$\chi_f$&$\chi_a$\cr 
 \hline
 0&0&0.0654&5.79&0.56&3.72&0.004&0.006\cr
 0.01&0.01&0.0661&5.79&0.60&3.81&0.005&0.006\cr
 -0.01&-0.01&0.0647&5.79&0.52&3.65&0.003&0.015 \cr
 0.01&-0.01&0.0791&5.67&0.40&3.31&0.002&0.048\cr
 -0.01&0.01&0.0541&5.90&1.46&4.72&0.017&0.056\cr
 -0.005&0.005&0.0595&5.85&0.77&4.07&0.007&0.031\cr
\hline
\end{tabular}
\caption{Values of $C$, $\beta_m$, $\cal A$, $\cal S $, $\chi_f$, and $\chi_a$ defined in the text for the dilogarithm model, corresponding to various choices of $\delta_9$ and $\delta_{10}$. All the fits refer to a comparison with the 26th order.}
\end{table}

\begin{table}[t]
\begin{tabular}{||c|c|c|c|c|c|c|c||}
\hline
$\delta_9$&$\delta_{10}$&$C_1$&$-d_1$&$\cal A$&$\cal S $ &$\chi_f$&$\chi_a$\cr 
 \hline
 0&0&0.219&3.82&0.63&3.90&0.004&0.08\cr
 0.01&0.01&0.221&3.82&0.70&4.01&0.006&0.07\cr
 -0.01&-0.01&0.216&3.82&0.57&3.81&0.003&0.08\cr
 0.01&-0.01&0.263&3.72&0.42&3.44&0.002&0.03\cr
 -0.01&0.01&0.182&3.92&2.29&5.24&0.024&0.12\cr
  -0.005&0.005&0.199&3.87&0.94&4.33&0.009&0.10\cr
\hline
\end{tabular}
\caption{Values of $C_1$, $-d_1$, $\cal A$, $\cal S $, $\chi_f$, and $\chi_a$ defined in the text for the integral model, corresponding to various choices of $\delta_9$ and $\delta_{10}$. All the fits refer to a comparison with the 26th order.}
\end{table}

In summary, we obtained values of $\cal S$ between 3.3 and 4.3 for fits with a reasonable $\chi_f$. The boundary of this interval corresponds to fits that have a larger $\chi_a$. These results seem to favor the idea that the NPP scales 
like $a^4$ over the idea that the NPP scales 
like $a^2$. However, no strong conclusions can be drawn because, the extrapolation models are empirical. 
The only indication of their approximate validity is that they provide consistent results up to some reasonably large order. It should also be pointed out that there is no universal agreement \cite{itep,grunberg97} 
about what should be done about the Landau pole in Eq. (\ref{eq:renormalon}). 
In order to draw strong conclusions, one would need either, numerical values of the coefficients up to significantly larger order than available now, 
or extrapolation models supported by a detailed analysis of graphs having significant contributions in {\it lattice} perturbation theory.

\section{Parametrization of the force scale}
\label{sec:force}

In the previous section, we considered the possibility of having the minimal error proportional to the fourth power of the two loop renormalization group invariant scale. 
What is attractive about this possibility is that it has the generic form 
$A\ \lambda^B \ {\rm e}^{-C/\lambda}$. However, the extrapolations suggest that it is not
a very accurate parametrization of the envelope of the accuracy curves and that a $(a/r_0)^4$ parametrization seems more promising. 
In this section, we consider  simple parametrizations of $a$.

\begin{figure}
\includegraphics[width=3.2in]{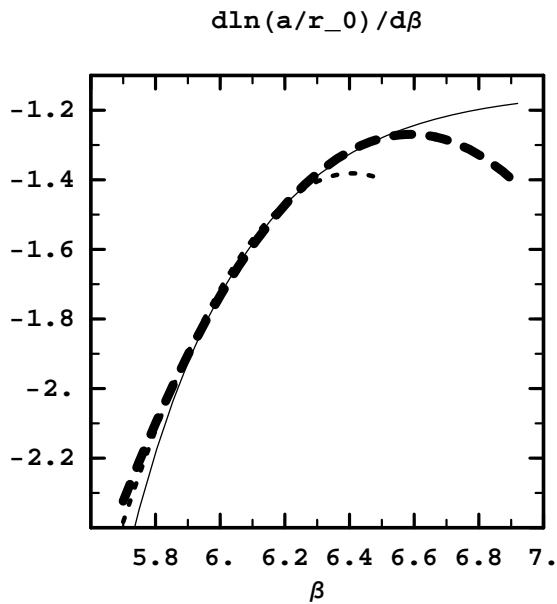}
\caption{$ d {\rm ln}(a/r_0)/d\beta$ using Ref. \cite{necco01} (thick dashes), \cite{guagnelli98} (small dashes) and Eq. (\ref{eq:fitforce}) (solid line).}
\label{fig:deforce}
\end{figure}

First we study the derivative of ln($a$) with respect to $\beta$ with the first two universal terms subtracted. It should be noted that the derivative emphasizes the cubic terms in the 
expansion of ln($a$) and the difference between the two expansions from Refs. \cite{guagnelli98} and \cite{necco01} is more pronounced than without the derivative (see 
Fig. \ref{fig:deforce}). 
We propose the following parametrization. 
\begin{equation}
d {\rm ln}(a/r_0)/d\beta=-(4\pi^2/33) + (51/121)\beta^{-1}+A_1{\rm e}^{-A_2\beta}\ .
\label{eq:fitforce}
\end{equation}
This form was originally motivated by our idea of obtaining a NPP that could be 
calculable semi-classically and was found later in good agreement with an independent argument \cite{allton96} 
(see below). We have also tried power parametrizations of the form $B_1\beta^{B_2}$ and found that it requires 
a large value of $B_2$ (close to 16) not very stable under small changes of the fiting interval.
Using the Taylor expansion of Ref. \cite{necco01} to produce a set of numerical values between 5.9 and 6.3, subtracting the first two terms of the r. h. s. and taking the log, we can determine numerically the parameters $A_1$ and $A_2$ from a linear fit. The result is  
$A_1=-1.35 \  10^{7}$ and $A_2=2.82$. 
For these values the exponential corrections become larger than the two universal terms for 
$\beta<5.8$. 
Integrating and fitting the constant of integration in the 
same interval, we found
\begin{equation}
{\rm ln}(a/r_0)=A_3-(4\pi^2/33)\beta + (51/121){\rm ln}(\beta)-(A_1/A_2){\rm e}^{-A_2\beta}\ .
\label{eq:som}
\end{equation}
with $A_3=4.5276$. This constant was obtained by plotting the l. h. s. minus all the other terms of the r. h. s.  versus $\beta$. A nice plateau appears between $\beta = 5.9$ and 6.3. The extremal values in this interval are 4.5272 and 4.5282. 

Note that the value of $A_2$ seems consistent with the idea \cite{allton96} 
of using $a_{pert.}^2$ corrections for this quantity. The symbol $a_{pert}$ refers to the 
one-loop or two-loop expression which in the short $\beta$ interval considered here 
can hardly be distinguished from each other. For this reason, we have not introduced any power correction in the last term of Eq. (\ref{eq:fitforce}). 
The assumption of $a_{pert.}^2$ corrections fixes $A_2=8\pi^2/33\simeq 2.4$ which is close to the value 2.82 obtained above. 

It is possible to use Eq. (\ref{eq:som}) to predict ${\rm ln}(a/r_0)$ at large $\beta$. 
For instance, at $\beta=7.5$, we obtain -3.59 (for -3.63 in Ref. \cite{guagnelli02}) and 
-4.74 at $\beta=8.5$ (for -4.81 in Ref. \cite{guagnelli02}).
It is also possible to obtain the lattice  scale $\Lambda_L$ from the constant of integration 
$A_3$, namely $\Lambda_L={\rm exp}(-A_3)/r_0\simeq 4.4\ MeV$.

\section{Conclusions}
We have considered the definition of the non-perturbative part of a quantity as the difference between its numerical value and the perturbative series truncated by dropping the order of minimal contribution (which is coupling dependent) and the higher orders. For the anharmonic oscillator, the double-well potential and the single plaquette gauge theory, the non-perturbative part can be parametrized as 
$A\ \lambda^B \ {\rm e}^{-C/\lambda}$ and the constants $A$, $B$ and $C$ can be calculated analytically. 

For lattice QCD in the quenched approximation, the perturbative series for the average 
plaquette is dominated at low order by a complex singularity and at this point, the asymptotic behavior can only be reached by using extrapolations. We have considered two extrapolations that provide a consistent description of the series up to order 20-25. 
More work is needed to understand these extrapolations better. A diagrammatic interpretation 
of the dilogarithm formula should be possible. The similarities between the 
one plaquette integral and the renormalon motivated integral suggest that it might be possible to 
derive the latter from an effective action for a single plaquette, obtained from the large 
volume partition function after integrating over all the other links. Ultimately, the two 
approximations should be put together using dispersion relations in the complex $\beta$ plane.

The two extrapolations favor the idea that the non-perturbative part is a power of the lattice spacing calculated using 
the force method and that the power is close to 4.
We found a parametrization of the force scale as the 
two loop universal terms with exponential corrections consistent with $a_{pert}^2$ corrections. 
These corrections become quite important when $\beta$ is near or below 6. 

For large $\beta$, we can treat the exponential corrections as small quantities and expand in 
power of these small quantities. The NPP of the plaquette is then 
written as superposition of terms of the form $A_j\ \beta^{B_j} \ {\rm e}^{-C_j\beta}$. 
If the constants $B_j$ and $C_j$ can be determined from general arguments, we only need to 
determine the amplitudes $A_j$, starting from those with smallest $C_j$. 
Calculating more terms of the perturbative series and 
calculating more accurate numerical values of the average plaquette at large $\beta$, 
would help in the empirical determination of these constants. 
As $\beta$ becomes large, the lattice spacing becomes small and the NPP is small. It is plausible that 
the amplitudes could be calculated by semi-classical methods in the continuum.
This is the most challenging part of the program. A method that comes to mind is the instanton 
calculus. However, the infrared sensitivity of the procedure is notorious and phenomenological input such as the gluon condensate is often invoked to produce realistic 
formulas \cite{shuryak98}. Having accurate empirical formulas for a well defined model 
(quenched) lattice QCD would help to revisit this question from scratch.

\begin{acknowledgments}
We acknowledge valuable discussions with C. Allton, M. Creutz, F. di Renzo, A. Duncan, G. Parisi, P. Rakow and G.C. Rossi. 
This 
research was supported in part  by the Department of Energy
under Contract No. FG02-91ER40664. 
\end{acknowledgments}

\end{document}